\documentclass[a4paper,11pt]{article}
\pdfoutput=1 

\usepackage{jheppub} 

\usepackage[T1]{fontenc} 
\usepackage{slashed}

\title{\boldmath Can electron and muon $g-2$ anomalies be jointly explained in SUSY ?}

\author[a,b]{Song Li,}
\author[a,b]{Yang Xiao,}
\author[a,b]{Jin Min Yang}

\affiliation[a]{CAS Key Laboratory of Theoretical Physics, Institute of Theoretical Physics, Chinese Academy of Sciences, Beijing 100190, China}
\affiliation[b]{School of Physics, University of Chinese Academy of Sciences,  Beijing 100049, China}

\emailAdd{lisong@itp.ac.cn}
\emailAdd{xiaoyang@itp.ac.cn}
\emailAdd{jmyang@itp.ac.cn}

\abstract{
The FNAL+BNL measurements for muon $g-2$ is $4.2\sigma$ above the SM prediction, and the Berkeley $^{133}$Cs measurement for the fine-structure constant $\alpha_{\rm em}$ leads to the SM prediction for electron $g-2$ which is $2.4\sigma$ above the experimental value. Hence, a joint explanation of both anomalies requires a positive contribution to muon $g-2$ and a negative contribution to electron $g-2$, which is rather challenging. In this work we explore the possibility of such a joint explanation in the minimal supersymmetric standard model (MSSM). Assuming no universality between smuon and selectron soft masses, we find out a part of parameter space for a joint explanation at $2\sigma$ level, i.e., $\mu M_1,\mu M_2<0$, $m_{L1}, m_{E2}<200$ GeV, $m_{L2}$ being much larger than the soft masses of other sleptons, $|M_1|<125$ GeV and $\mu<400$ GeV. This part of parameter space can survive LHC and LEP constraints, but gives an over-abundance for dark matter if the bino-like lightest neutralino is assumed to be the dark matter candidate. With the assumption that the dark matter candidate is a superWIMP (say a pseudo-goldstino in multi-sector SUSY breaking scenarios, whose mass can be as light as GeV and produced from the late-decay of the thermally freeze-out lightest neutralino), the dark matter problem can be avoided.  So, we conclude that the MSSM may give a joint explanation for the muon and electron $g-2$ anomalies at $2\sigma$ level (the muon $g-2$ anomaly can be even ameliorated to $1\sigma$).            
}

\begin{document} 
\maketitle
\flushbottom

\section{Introduction}
\label{sec:intro}

There has been a long-standing discrepancy between the standard model (SM) prediction and experiment for muon anomalous magnetic moment $a_\mu=(g-2)_\mu/2$. The combined result of the FNAL E989 experiment   ~\cite{PhysRevLett.126.141801} and the BNL experiment gives a value 
which is $4.2\sigma$ above the SM prediction ~\cite{Aoyama:2020ynm}:
\begin{equation} \label{eq:delta-amu}
    \begin{aligned}
        \Delta a_{\mu}^{\rm{Exp-SM}} &=a_\mu^{\rm Exp}-a_\mu^{\rm SM} \\
        &= (2.51\pm 0.59) \times 10^{-9}.
    \end{aligned}
\end{equation}

On the other hand, for electron anomalous magnetic moment $a_e$, the SM predicted value~\cite{Aoyama:2019ryr} 
derived from the measurement of the fine-structure constant $\alpha_{\rm em}$ using $^{133}$Cs atoms at Berkeley~\cite{Parker:2018vye} is $2.4\sigma$ above the  electron $g-2$ experimental value~\cite{Hanneke:2008tm}:
\begin{equation} \label{eq:delta-ae}
    \begin{aligned}
        \Delta a_e^{\rm{Exp-SM}} &=a_e^{\rm Exp}-a_e^{\rm SM}(\text{Cs}) \\
        &=(-8.8\pm 3.6)\times 10^{-13}.
    \end{aligned}
\end{equation}
However, another experimental result of $\alpha_{\rm em}$ measured with $^{87}$Rb atoms at Laboratoire Kastler Brossel (LKB)~\cite{Morel:2020dww} gives a value of $a_e$~\cite{Aoyama:2012wj} which agrees with the electron $g-2$ experimental value~\cite{Hanneke:2008tm}. 
So far, the cause of the discrepancy between the Berkeley and LKB results is not clear. Obviously, if the Berkeley result is correct, it may serve as a plausible hint for new physics; if the LKB result is correct, there will be no need for new physics to provide a contribution to $(g-2)_e$. 

Actually, neither the Berkeley result for electron $g-2$ nor the FNAL+BNL result for muon $g-2$ 
can serve as a robust evidence for new physics beyond the SM. Whereas, while waiting for more forthcoming independent experiments to make confirmation, many theorists have chosen to keep open minds and intensively studied the implications of these anomalies for new physics.  
In this work, we keep open minds and study the implications of these anomalies for low energy supersymmetry.
 
The $g-2$ of a charged lepton relates to the new physics scale $\Lambda$ as~\cite{book:Jegerlehner:2017}
\begin{equation}\label{eq:delta-alepton}
    \frac{\delta a_{\ell}}{a_{\ell}}\propto \left(\frac{m_{\ell}}{\Lambda}\right)^2,
\end{equation}
where $\Lambda$ is assumed to be much larger than the lepton mass $m_{\ell}$.  If the new physics is the same for different lepton flavors, we may expect 
\begin{equation}
    \frac{\delta a_e}{\delta a_\mu} \approx \left(\frac{m_e}{m_\mu}\right)^2\approx \frac{1}{43000}.
\end{equation}%
But from eq.\eqref{eq:delta-amu} and eq.\eqref{eq:delta-ae} we obtain
\begin{equation}\label{eq:ratioofMDM}
    \frac{\Delta a_e^{\rm{Exp-SM}} }{\Delta a_\mu^{\rm{Exp-SM}} }\approx -15\times \frac{1}{43000},
\end{equation}%
which implies that the new physics should not be flavor blind \cite{Giudice:2012ms}. This brings us a serious challenge for building new physics models. Intuitively, in order to jointly explain the electron and muon $g-2$ anomalies, new physics models must have different couplings for charged leptons of different flavors. This will suggest us to consider some new physics models with lepton flavor universality violation. However, in some models the manifest breaking of lepton flavor universality is not required~\cite{Hiller:2019mou,Hiller:2020fbu,Bissmann:2020lge}.

There have been many studies trying to explain  electron and muon $g-2$ jointly:
(i) Using the two-Higgs-doublet model (2HDM) or its extended versions \cite{Rose:2020nxm,Botella:2020xzf,Hernandez:2021tii,Jana:2020pxx,Keung:2021rps,Li:2020dbg,Han:2018znu},
among which the aligned 2HDM with right-handed neutrinos was used in   
~\cite{Rose:2020nxm}, the 2-loop contribution in 2HDM with flavor conservation was considered in ~\cite{Botella:2020xzf}, while a neutral scalar $H$ with mass satisfying $\mathcal{O}(1)\,\text{MeV}<m(H)<\mathcal{O}(1)\,\text{GeV}$ was used in ~\cite{Jana:2020pxx};
(ii) Using leptoquarks ~\cite{Dorsner:2020aaz,Keung:2021rps,Cornella:2019uxs} and axion-like particles \cite{Cornella:2019uxs};
(iii) Using flavor models ~\cite{Hernandez:2021tii,Calibbi:2020emz} which attempted to solve the fermion mass hierarchy problem while provide a joint explanation for the muon/electron $g-2$ anomalies;
(iv) Using the inverse type-III seesaw model with a pair of vector-like leptons ~\cite{Escribano:2021css}; 
(v) Using an abelian flavor symmetry \cite{Crivellin:2018qmi} to make electron and muon decouple to circumvent the MEG bound
of $\mathcal{B}(\mu^+\to e^+\gamma)$ ~\cite{TheMEG:2016wtm} (note that $Z'$-model cannot explain the muon/electron $g-2$ anomalies under the MEG bound, as shown in ~\cite{CarcamoHernandez:2019ydc,Bodas:2021fsy});
(vi) Using a flavor-dependent global $U(1)$ symmetry and a discrete $\mathbb{Z}_2$ symmetry with some additional fermion and scalar fields~\cite{Chen:2020tfr}. 

As a leading candidate for new physics models, the MSSM (minimal supersymmetric standard model) was also considered ~\cite{Dutta:2018fge} to interpret $\Delta a_{e,\,\mu}$ while escaping the $\mathcal{B}(\mu^+\to e^+\gamma)$ constraint by assuming 1-3 flavor violation. 
Another attempt in the MSSM without assumption of flavor violation but with flavor non-universality was given in ~\cite{Badziak:2019gaf}, where the authors decouple the right-hand smuon and choose a negative soft mass parameter $M_1$ to achieve the goal. Furthermore, one can explain $\Delta a_{e,\,\mu}$ jointly by using the SUSY threshold correction to change the selectron and smuon Yukawa couplings including size and sign~\cite{Endo:2019bcj}. In addition, the B-L SSM and an extended NMSSM (next-to-minimal supersymmetric standard model) was used for a joint explanation of muon/electron $g-2$ anomalies~\cite{Yang:2020bmh,Cao:2021lmj}. 

Anyway, it is rather challenging for the MSSM to simultaneously explain electron and muon $g-2$. Although previous studies found out a plausible part of parameter space in the MSSM, some relevant constraints were not considered (say from dark matter) or not fully considered (say from collider experiments).  Given the popularity of the MSSM and the plausible new physics hints from the muon/electron $g-2$ anomalies,  we in this work revisit the MSSM to give a more comprehensive study. We will explore the MSSM parameter space to figure out the possibility to accommodate  the muon/electron $g-2$ anomalies under other relevant constraints from collider experiments and dark matter measurements. 

This work is organized as follows. In Sec.~\ref{sec:GM2inMSSM}, we describe the MSSM contributions to muon/electron $g-2$.  In Sec.~\ref{sec:parameter-area}, we explore parameter space to accommodate 
the muon/electron $g-2$ anomalies.  
In Sec.~\ref{sec:DM-and-LHC}, we show relevant constraints on the favored parameter space. Finally, we conclude in Sec.~\ref{sec:conclusions}.

\section{MSSM contributions to electron and muon \texorpdfstring{$g-2$}{g-2}}
\label{sec:GM2inMSSM}

The MSSM contributions to a charged lepton $\ell$ (electron or muon) $g-2$ mainly come from neutralino-slepton and chargino-sneutrino loops. The analytical expressions contributed by these two parts are given by~\cite{Martin:2001st}
\begin{align}\label{eq:leptonMDM1}
	\delta a_{\ell}^{\chi^0} = \frac{m_{\ell}}{16\pi^2}
    \sum_{i,m}&\left\{ -\frac{m_{\ell}}{ 12 m^2_{\tilde\ell_m}}
    (|n^L_{im}|^2+ |n^R_{im}|^2)F^N_1(x_{im}) \right.\notag\\
    &\quad\left.+\frac{m_{\chi^0_i}}{3 m^2_{\tilde\ell_m}}
    {\rm Re}[n^L_{im}n^R_{im}] F^N_2(x_{im})\right\}, \\
    \label{eq:leptonMDM2}
    \delta a_{\ell}^{\chi^\pm} = \frac{m_{\ell}}{16\pi^2}\sum_k
    &\Bigg\{ \frac{m_\ell}{ 12 m^2_{\tilde\nu_\ell}}
    (|c^L_k|^2+ |c^R_k|^2)F^C_1(x_k) \notag\\
    &\left.\quad +\frac{2m_{\chi^\pm_k}}{3m^2_{\tilde\nu_\ell}} {\rm Re}[ c^L_kc^R_k] F^C_2(x_k)\right\},
\end{align}
where $x_{im}=m^2_{\chi^0_i}/m^2_{\tilde l_m}$ and $x_k=m^2_{\chi^{\pm}_k}/m^2_{\tilde \nu_l}$. The definitions of $n^{L,R}$, $c^{L,R}$ and $F_{1,2}^{C,N}$ can be found in appendix. 

In this work, we mainly use eq.~\eqref{eq:leptonMDM1} and eq.~\eqref{eq:leptonMDM2} to calculate the muon/electron $g-2$ and some significant 2-loop corrections will also be included. After considering the 2-loop corrections, we have 
\begin{equation}\label{eq:MDM2loop}
   \begin{aligned}
       \delta a_\ell^{\rm SUSY}=&\left(1-\frac{4\alpha}{\pi}\ln\frac{M_{\rm SUSY}}{m_\ell}\right)\times \\
       &\quad\left(\frac{1}{1+\Delta_\ell}\right)\delta a_\ell^{\rm SUSY,\,1L},
   \end{aligned}
\end{equation}
where $\delta a_\ell^{\rm SUSY,\,1L}$ denotes one-loop SUSY contributions.  
The term in the first bracket arises from the leading-logarithmic QED correction~\cite{Degrassi:1998es}. This correction takes into account  renormalization group evolution of effective operators from $M_{\rm SUSY}$ to $m_\ell$ scale, which can lead to a reduction of about $7\%$ for $\delta a_\mu$, and a reduction of about $11\%$ for $\delta a_e$. The term in the second bracket of eq.~\eqref{eq:MDM2loop} arises from the $\tan\beta$-enhanced loop diagrams that can correct the Yukawa couplings of sleptons, and a resummation has been made~\cite{Marchetti:2008hw,Carena:1999py}. $\Delta_\ell$ is given by 
\begin{equation}\label{eq:Deltal}
   \begin{split}
       \Delta_\ell=&-\mu\tan\beta\,\frac{g_2^2 M_2}{16\pi^2}\left[I(m^2_{\chi^{\pm}_1},m^2_{\chi^{\pm}_2},m_{\tilde \nu_\ell}^2)+\right.\\
       & \qquad \left.\frac12 I(m^2_{\chi^{\pm}_1},m^2_{\chi^{\pm}_2},m_{\tilde \ell_L}^2)\right]-\\
       &\quad \mu\tan\beta\,\frac{g_1^2 M_1}{16\pi^2} \left[I(\mu^2,M_1^2,m_{\tilde \ell_R}^2)-\right.\\
       & \qquad \left.\frac12 I(\mu^2,M_1^2,m_{\tilde \ell_L}^2)-I(M_1^2,m_{\tilde \ell_L}^2,m_{\tilde \ell_R}^2)\right],
   \end{split}
\end{equation}
where the loop function $I(a,b,c)$ is given by 
\begin{equation}
       I(a,b,c)=-\frac{ab\ln(a/b) + bc\ln(b/c) + ca\ln(c/a)}{(a-b)(b-c)(c-a)}.
\end{equation}
We use some tricks to prevent enormous errors while avoid false singularities of $I(a,b,c)$. We note that $I(a,b,c)$ is fully symmetric to the three parameters, and therefore  we assume $a\leq b\leq c$ and define $x=a/b$, $y=c/b$. Then we obtain
\begin{equation}
       I(a,b,c)=\frac{1}{c-a}\left(y\times \frac{\ln y}{y-1}-x\times\frac{\ln x}{x-1}\right).
\end{equation}
If $x\approx 1$ and $a~\slashed{\approx}~c$, one can perform Fourier expansion for $\ln x/(x-1)$ around $x=1$ for the calculation. The practice is similar when $y\approx 1$ and $a~\slashed{\approx}~c$. If $a\approx c$, we demand the Fourier expansion of the binary function. Defining $\delta x=x-1$ and $\delta y=y-1$, then we have 
\begin{equation}
   \begin{split}
       I(a,b,c)=\frac{1}{2b}& \left[ 1-\frac{\delta x+\delta y}{3}+\frac{g(\delta x,\,\delta y,\,2)}{6}-\right. \\
       & \left. \frac{g(\delta x,\,\delta y,\,3)}{10}+\frac{g(\delta x,\,\delta y,\,4)}{15}+\cdots\right],
   \end{split}
\end{equation}
where
\begin{equation}
       g(\delta x,\,\delta y,\,n)=\sum_{i=0}^n\delta x^i\delta y^{n-i}.
\end{equation}

From eq.~\eqref{eq:Deltal} we know that this correction can be enhanced by a large $\mu\tan\beta$.  We calculate $\delta a_{e,\,\mu}^{\rm SUSY}$ using the above results. Our code is cross-checked with \textbf{CPsuperH 2.3}~\cite{Lee:2003nta,Lee:2007gn,Lee:2012wa}, and the difference between numerical results is less than $0.1\%$ in magnitude. 

In order to find out the parameter space that can jointly explain $\Delta a_{e,\,\mu}^{\rm{Exp-SM}}$, we classify the SUSY contributions approximately as ~\cite{Moroi:1995yh,Stockinger:2006zn} 
\begin{eqnarray}
       \delta a_\ell (\tilde{W},\tilde{H},\tilde{\nu}_\ell)&\simeq& 15\times 10^{-9}R\left(\frac{(100 {\rm GeV})^2}{\mu~M_2}\right),~  \label{a_l_loop_1}\\
       \delta a_\ell(\tilde{W},\tilde{H},\tilde{\ell}_L)&\simeq& -2.5\times 10^{-9}R\left(\frac{(100 {\rm GeV})^2}{\mu~ M_2}\right),~  \label{a_l_loop_2}\\
       \delta a_\ell(\tilde{B},\tilde{H},\tilde{\ell}_L)&\simeq& 0.76\times 10^{-9}R\left(\frac{(100 {\rm GeV})^2}{\mu~ M_1}\right),~ \label{a_l_loop_3}\\
       \delta a_\ell(\tilde{B},\tilde{H},\tilde{\ell}_R)&\simeq& -1.5\times 10^{-9}R\left(\frac{(100 {\rm GeV})^2}{\mu~M_1}\right),~ \label{a_l_loop_4}\\
       \delta a_\ell(\tilde{\ell}_L,\tilde{\ell}_R,\tilde{B})&\simeq& 1.5\times 10^{-9}R\left(\frac{(100 {\rm GeV})^2(\mu M_1)}{m_{\tilde{\ell}_L}^2m_{\tilde{\ell}_R}^2}\right),\quad~  \label{a_l_loop_5}
\end{eqnarray}
where $\ell=e,\,\mu$, and $R=(m_\ell/m_\mu)^2(\tan\beta/10)$ for brevity. Eqs.(\ref{a_l_loop_1}-\ref{a_l_loop_5}) are obtained through simplification under certain conditions, focusing on the order of magnitude. $\delta a_{\ell}$ is derived from the loop correction. As the masses of the sleptons in the loops increase, the contribution of the loop diagrams becomes smaller. This dependency does not appear explicitly in eqs.(\ref{a_l_loop_1}-\ref{a_l_loop_4}). 
We see that the MSSM contribution to $g-2$ of a charged lepton can be positive or negative. To have a negative $\delta a_e^{\rm{SUSY}}$ and a positive $\delta a_\mu^{\rm{SUSY}}$, we need to assume non-universality between smuon and selectron soft masses. 
We can find out two typical scenarios to have a negative $\delta a_e^{\rm{SUSY}}$ and a positive $\delta a_\mu^{\rm{SUSY}}$ for a joint explanation of muon/electron $g-2$ anomalies:
\begin{itemize}
\item[(i)]  Use the chargino-sneutrino loop in eq.(\ref{a_l_loop_1}) to give a positive $\delta a_\mu^{\rm{SUSY}}$ assuming $\mu M_2>0$; use the bino-selectron loop in eq.(\ref{a_l_loop_5}) to give a negative $\delta a_e^{\rm{SUSY}}$ assuming $\mu M_1<0$. 
This scenario was studied in ~\cite{Badziak:2019gaf} and will not be restudied in this work.
\item[(ii)]  Use the chargino-sneutrino loop in eq.(\ref{a_l_loop_1}) to give a 
negative $\delta a_e^{\rm{SUSY}}$ assuming $\mu M_2<0$;  use the bino-higgsino-smuon loop in eq.(\ref{a_l_loop_4}) to give a positive $\delta a_\mu^{\rm{SUSY}}$ assuming $\mu M_1<0$.
This scenarios will be studied in detail in this work.
\end{itemize}
We would like to comment on the virtues of the second scenario. 
Since the SUSY contributions to $a_e$ are suppressed by $(m_e/m_\mu)^2$ compared with the contrubutions to $a_\mu$, we need much larger SUSY loop effects for $a_e$.   
From the above formulas we see that the coefficient of 
the chargino-sneutrino loop in eq.(\ref{a_l_loop_1}) 
is one order of magnitude higher than other four kinds of loops. 
Using it with assumption $\mu M_2<0$ to explain $\Delta a_e^{\rm{Exp-SM}}$ in Eq.~\eqref{eq:ratioofMDM} is a wise choice. 
In this way, winos and higgsinos do not need to be very light. 
On the other hand, in order for $\delta a_\mu^{\rm{SUSY}}$ not to get a sizable negative 
contribution from the chargino-sneutrino loop in eq.(\ref{a_l_loop_1}),  
we assume the left-handed smuon $\tilde{\mu}_L$ to be very heavy (note that $\tilde{\nu}_\mu$ and $\tilde{\mu}_L$ are in a $SU(2)_L$ doublet and thus have the same soft mass).  Therefore, in this scenario only the bino/higgsino-smuon loop in eq.(\ref{a_l_loop_4}) is left to give a positive $\delta a_\mu^{\rm{SUSY}}$ assuming $\mu M_1<0$.
 In this way, the right-handed smuon, bino and higgsinos are required to be light. As will be shown in the following, there is indeed a parameter space for a joint explanation of muon/electron $g-2$ anomalies, which can survive collider constraints. However, since the lightest neutralino is bino-like, its thermal freeze-out number density is large; thus it cannot be the dark matter candidate and must decay (to a lighter stable particle like a gravitino or speudo-goldstino as the dark matter particle) after thermal freeze-out.    
 
\section{MSSM parameter space for a joint explanation}
\label{sec:parameter-area}
According to previous discussions, we require $\mu M_1<0$ and $\mu M_2<0$. In this work we assume that the parameters are all real to avoid CP violation. We can freely choose $\mu>0$, $M_1<0$ and $M_2<0$ (similar results can be obtained for the case of $\mu<0$). For simplicity, we choose $M_2$ and $\mu$ as the scan variables, and 
\begin{equation} \label{eq:MLandME}
       m_{L1}=m_{E2}=\min(|\mu|,\,|M_1|,\,|M_2|)+30\,\text{GeV},
\end{equation}
where the 30 GeV increment is to ensure that the slepton masses are above the LEP bound ~\cite{LEP:sleptons} (in our scan the minimal value of $|\mu|,\,|M_1|,\,|M_2|$ is 80 GeV) and the difference from $m_{\chi_1^0}$ is larger than 30 GeV, which helps to avoid the LHC search constraints for the compressed slepton-neutralino spartciles (see Fig.16 in ~\cite{ATLAS:2019lng}).  The collider constraints will be discussed in the proceeding section. 
In order to suppress $\delta a_\mu^{\rm{SUSY}}(\tilde{W},\tilde{H},\tilde{\nu}_\mu)$, we take $m_{L2 }=10m_{L1}$, with $L1$ and $L2$ denoting respectively the first and second generation left-handed sleptons (similarly $E2$ denotes the second generation righ-handed slepton, i.e., the  righ-handed smuon).  
Because $\delta a_e^{\rm{SUSY}}$ is dominated by $\delta a_e^{\rm{SUSY}}(\tilde{W},\tilde{H},\tilde{\nu}_e)$, the right-handed selectron mass $m_{R1}$ will not observably affect this part as long as it does not make $\delta a_e^{\rm{SUSY}}(\tilde{B},\tilde{H},\tilde{e}_R)$ too large. Hence we set $m_{R1}=5 m_{L1}$. 

\begin{figure*}[tbp]
\centering 
\includegraphics[width=.4\textwidth]{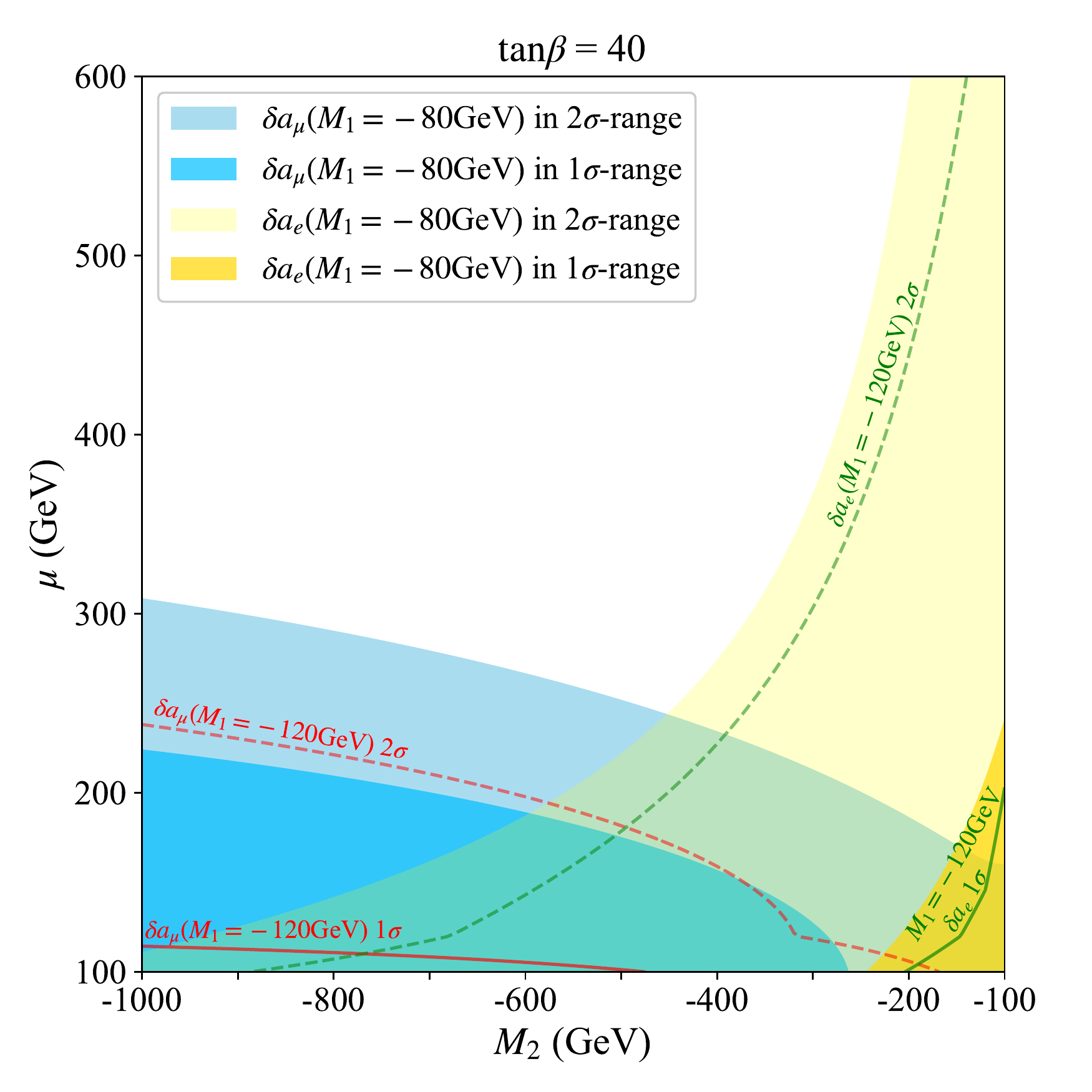}
\includegraphics[width=.4\textwidth]{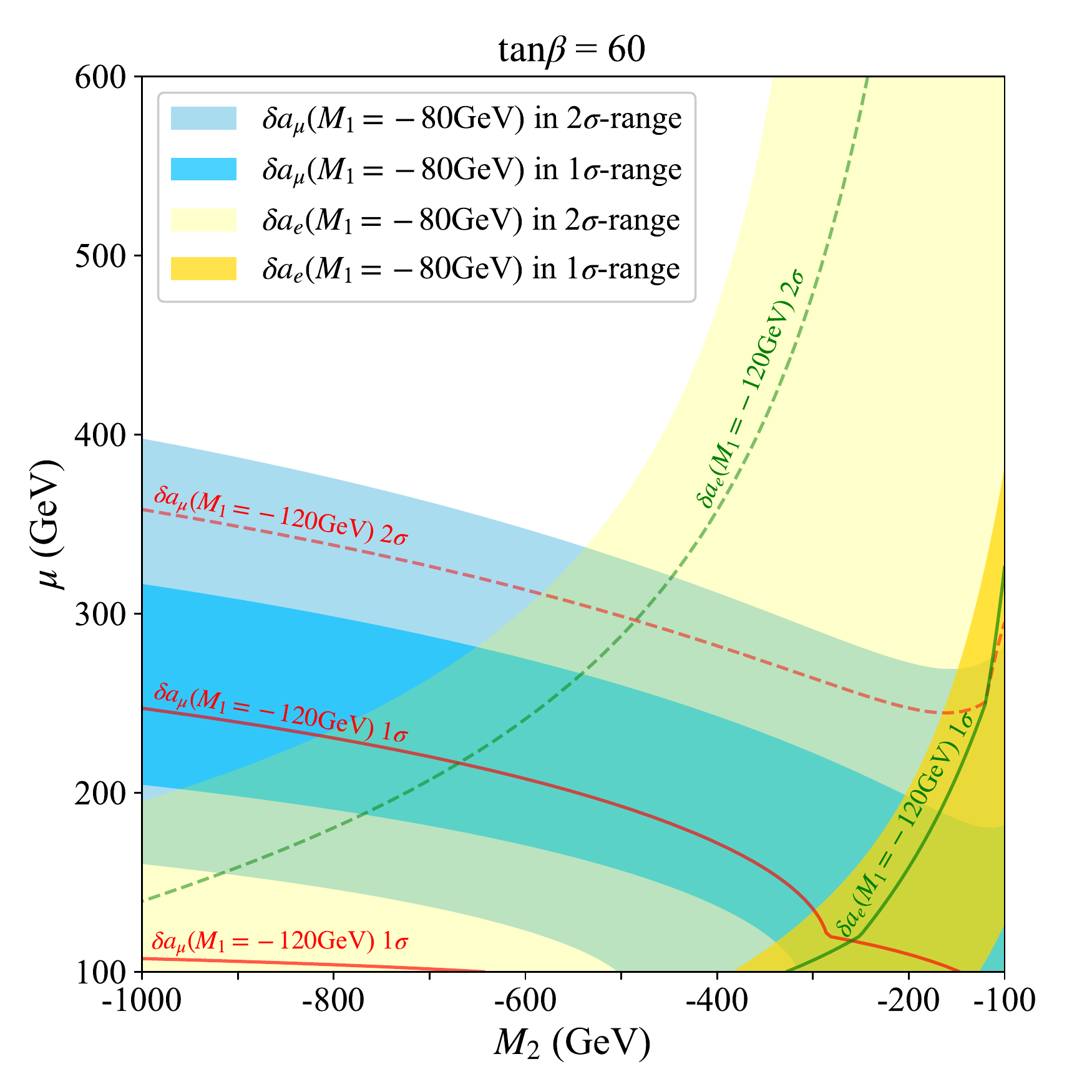}
\caption{\label{fig:onlyGM2} 
Contours of $\delta a_{e}^{\rm SUSY}$ and $\delta a_{\mu}^{\rm SUSY}$ on the plane of $\mu$ versus $M_2$ for $\tan\beta=40$ (left) and $\tan\beta=60$ (right). The colored regions correspond to $M_1=-80\text{~GeV}$ while the colored curves correspond to $M_1=-120~\text{GeV}$.  The $1\sigma$ and $2\sigma$ mean to explain $\Delta a_{e,\mu}^{\rm{Exp-SM}}$ at $1\sigma$ and $2\sigma$ levels, respectively. }
\end{figure*}

The results from our exploration of parameter space are shown in Fig.~\ref{fig:onlyGM2}. Because $\delta a_\mu^{\rm{SUSY}}(\tilde{B},\tilde{H},\tilde{\mu}_R)$ is the dominant contribution to $\delta a_\mu^{\rm{SUSY}}$, the value of $\delta a_\mu^{\rm{SUSY}}$ 
(especially its $1\sigma$ range required by the explanation of  $\Delta a_\mu^{\rm{Exp-SM}}$ ) is sensitive to $M_1$. 
With respect to the case with $|M_1|=80~\text{GeV}$, the $1\sigma$ range of $|M_1|=120~\text{GeV}$ moves down significantly. For $|M_1|>120~\text{GeV}$, the value of $\delta a_\mu^{\rm{SUSY}}$ will decrease rapidly, which is not shown in the figures.  As for $\delta a_e^{\rm{SUSY}}$, the dominant loop contribution is not sensitive to $M_1$. Hence even if $|M_1|$ is very large, $\delta a_e^{\rm{SUSY}}$ does not change drastically. Therefore, in order to satisfy the experimental value of $a_\mu$, $|M_1|$ cannot be greater than $\sim120$ GeV. This makes the tree-level mass of $\chi_1^0$ lower than 120 GeV.
   
From Fig.~\ref{fig:onlyGM2} we can also see that $\mu$ and $M_2$ have an inverse relationship when $\delta a_e^{\rm{SUSY}}$ has a fixed value. This is easy to understand because of the appearance  of $\mu M_2$ in eq.(\ref{a_l_loop_1}).   
However, $\delta a_\mu^{\rm{SUSY}}$ unexpectedly increases with the increase of $|M_2|$.
This is because $\delta a_\mu^{\rm{SUSY}}(\tilde{W},\tilde{H},\tilde{\nu}_\mu)$ is not suppressed enough by assuming $m_{L2}=10m_{L1}$, which is getting suppressed by increasing $|\mu M_2|$. If we set the value of $m_{L2}$ bigger than $10m_{L1}$, we can further reduce the dependence of $\delta a_\mu^{\rm{SUSY}}$ on $M_2$. In that way, the range of $M_2$ allowed by the experimental constraints will be extended.
   
From  Fig.~\ref{fig:onlyGM2} we note that for $\tan\beta=40$ the two $1\sigma$-ranges for the explanations of $\Delta a_{e,\mu}^{\rm{Exp-SM}}$ do not overlap. Therefore, a joint explanation of $\Delta a_{e,\mu}^{\rm{Exp-SM}}$ at $1\sigma$ level needs a larger $\tan\beta$. 
For $\tan\beta=60$ the two $1\sigma$-ranges of $\Delta a_{e,\mu}^{\rm{Exp-SM}}$ overlap, which, however, requires $\mu<200~\text{GeV}$ and $|M_2|<300~\text{GeV}$.
For a joint explanation of $\Delta a_{e,\mu}^{\rm{Exp-SM}}$ at $2\sigma$ level, the ranges of the relevant parameters
are relaxed significantly. 

\section{Dark matter and collider constraints}
\label{sec:DM-and-LHC}

\subsection{Dark matter constraints}
 We use \textbf{FlexibleSUSY 2.6.0}~\cite{Athron:2014yba,Athron:2017fvs} to calculate the mass spectrum of supersymmetric particles, and then use \textbf{MicrOMEGAs 5.2.7}~\cite{Belanger:2006is,Belanger:2008sj,Belanger:2010pz,Belanger:2013oya} to calculate the dark matter relic density $\Omega h^2$ and the LSP-nucleon scattering cross section. For the parameters that are not directly related to $\delta a_{e,\,\mu}^{\rm{SUSY}}$, we generally take $2~\text{TeV}$. In addition, the mass parameters of stau are taken as $10m_{L1}$ and the $A$-parameters that describe the trilinear soft-breaking terms are set to be 0.

\begin{figure*}[tbp]
\centering 
\includegraphics[width=.4\textwidth]{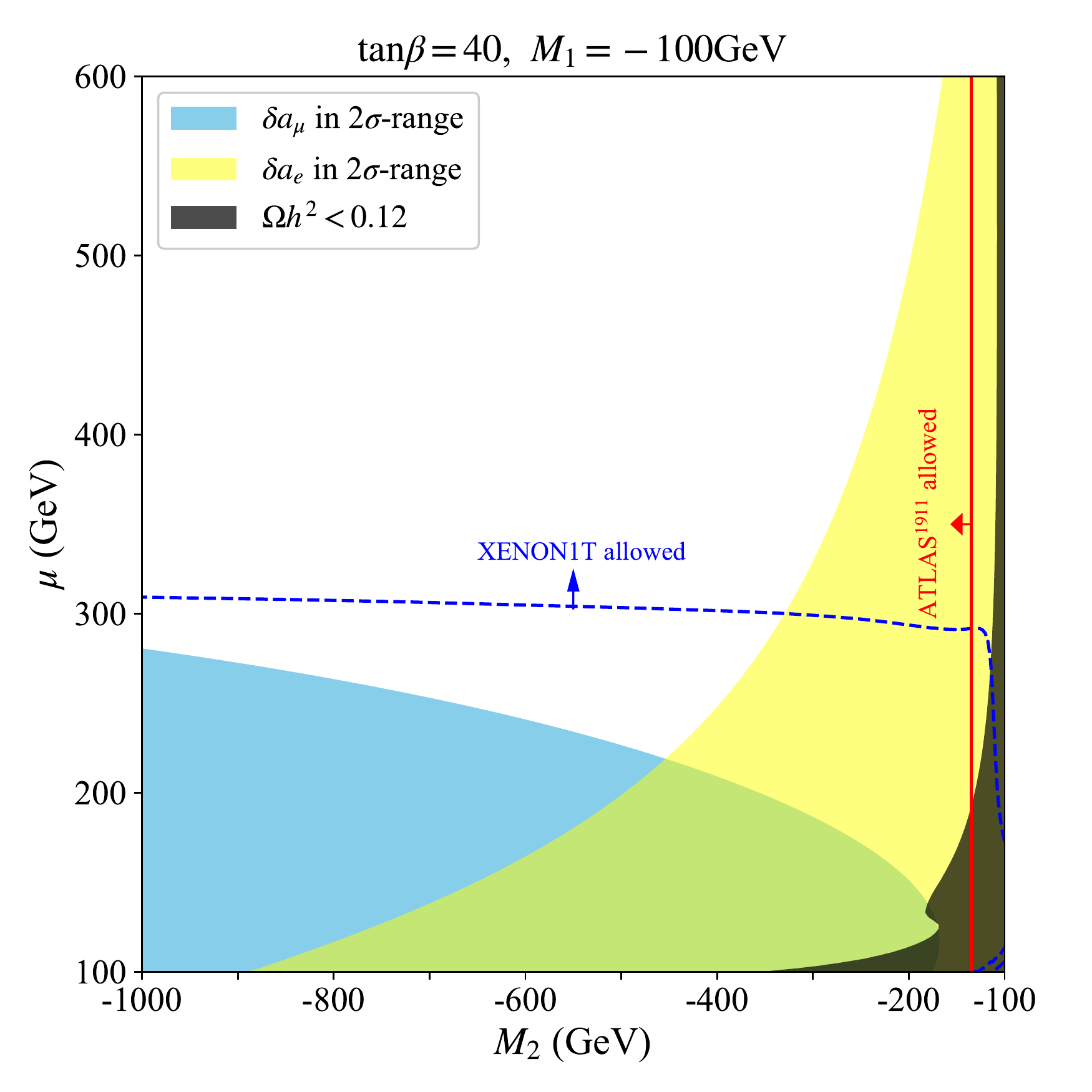}
\includegraphics[width=.4\textwidth]{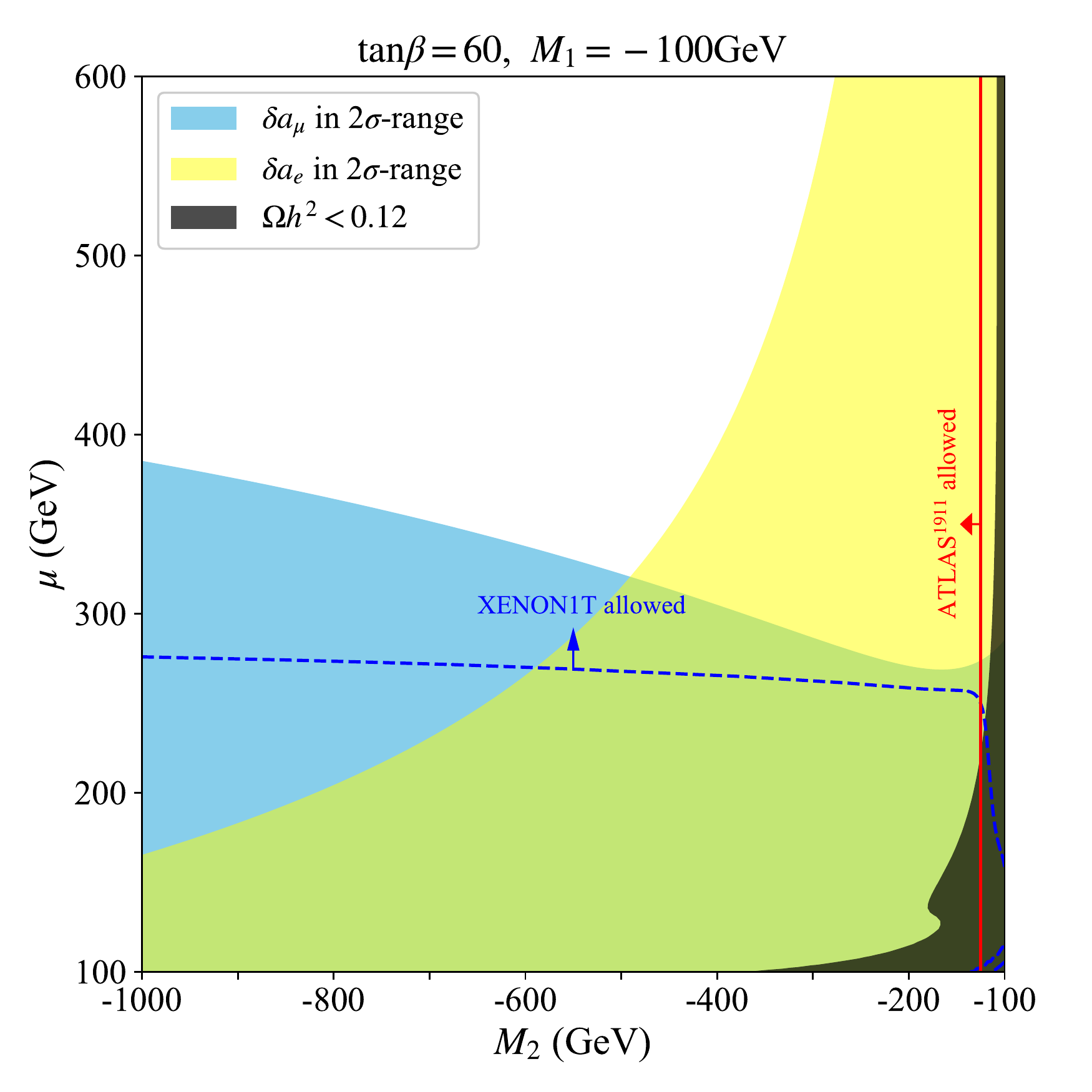}
\caption{\label{fig:DMandGM2}The parameter regions survived the dark matter relic density constraint and the XENON1T exclusion limits on the plane of $\mu$ versus $M_2$ with $M_1=-100~\text{GeV}$ and $\tan\beta=40,60$. The $2\sigma$ regions for explaining the muon and electron $g-2$ anomalies are also shown. For the curves of limits, the regions indicated by the arrows are the allowed  regions.}
\end{figure*}

We first assume the lightest neutralino  $\chi_0^1$ is the dark matter candidate and search for the parameter space that meets the following three requirements:
\begin{enumerate}
    \item The SUSY contributions $\delta a_{e,\,\mu}^{\rm{SUSY}}$ in the $2\sigma$ ranges of $\Delta a_{e,\,\mu}^{\rm{Exp-SM}}$;
    \item The dark matter thermal freeze-out relic density under the upper bound, i.e., $\Omega_{\chi_1^0 }h^2<0.12$~\cite{Planck:2018vyg};
    \item The XENON1T limits from the dark matter direct detection ~\cite{XENON:2017vdw}.
\end{enumerate}%
We scan over the parameter space and the results are shown in Fig.~\ref{fig:DMandGM2}. 
In the regions shown in this figure, the dominant component of the lightest neutralino is bino, whose thermal freeze-out 
relic density can easily give an over-abundance.  
For the bino-like dark matter to give a correct relic density, higgsinos or wino must be mixed into it,
which, however, will be not allowed by the XENON1T limits. 
As shown in this figure, only a very large $\tan\beta=60$ can give a corner of parameter space to satisfy both the dark matter 
relic density and the XENON1T limits, which, however, is not allowed by the LHC constraints~\cite{ATLAS:2019lng}.

So we conclude that for the SUSY contributions $\delta a_{e,\,\mu}^{\rm{SUSY}}$ in the $2\sigma$ ranges of $\Delta a_{e,\,\mu}^{\rm{Exp-SM}}$, the assumption of the lightest neutralino as the dark matter candidate cannot satisfy the dark matter constraints under the LHC search bounds. Note that in the framework of SUSY, the lightest neutralino is not the only candidate for cosmic dark matter. Instead, the lightest super particle (LSP) as the dark matter candidate can be a superWIMP (super-weakly interacting massive particle) like the gravitino ~\cite{Feng:2004gg,Wang:2004ib} (its goldstino component has a relatively stronger interaction than gravity) or pseudo-goldstino ~\cite{Argurio:2011hs} in multi-sector SUSY breaking with gauge mediation. 
As discussed in detail in ~\cite{Feng:2004gg}, in such superWIMP dark matter scenarios, 
 the superWIMPs produced thermally before inflation are diluted by inflation and the superWIMP dark matter is produced from the late decay of the lightest neutralinos which are produced from the thermal freeze-out after reheating (the reheating temperature is not high enough to thermally produce superWIMPs). Since the decay is one neutralino $\chi_0^1$ to one superWIMP $\tilde{G}$, i.e., $\chi_0^1\to \tilde{G}+X$ ($X=\gamma,Z,h$), 
 the superWIMP dark matter inherits the number density of the parent neutralinos and 
 hence its relic density is suppressed by a factor $m_{\tilde{G}}/m_{\chi^0_1}$, where 
 $m_{\chi^0_1}$ is $\mathcal{O}(100)~\text{GeV}$ and $m_{\tilde{G}}$ can be lighter than 
 $\mathcal{O}(1)~\text{GeV}$~\cite{Argurio:2011hs,Dai:2021eah}.
 So the relic density upper bound can be easily satisfied by such light superWIMP dark matter. 
 Of course, a superWIMP scatters with a nucleon super-weakly and the direct detection limits can also 
 be easily satisfied. 
 
 A possible problem caused by such superWIMP dark matter produced from the late decay of the lightest neutralinos is that the decay may release much energy to affect BBN if the decay happens after BBN.
 Such a problem and its constraints have been discussed in detail in ~\cite{Feng:2004gg,Cyburt:2002uv}. 
 Recently, it was found \cite{Gu:2020ozv} that late decay of the freeze-out neutralinos to very light gravitino dark matter can ameliorate the tension of Hubble constant. At a future lepton collider the decay of a bino-like neutralino to gravitino plus a photon may be tested \cite{Chen:2021omv}.    
 
\subsection{LHC constraints}
Now we consider constraints from colliders. We use \textbf{SPheno 4.0.4}~\cite{Porod:2003um,Porod:2011nf} to calculate the decay branching ratios of the relevant super particles. 
We consider two cases: 
\begin{eqnarray}
   \text{case 1} &\left\{ \begin{aligned} & M_1=-100~\text{GeV},~\tan\beta=40,\\
     & m_{L1}=m_{E2}=\frac{m_{E1}}{5}=\frac{m_{L2}}{10}=130~\text{GeV};
     \end{aligned}\right.
   \label{eq:ML130}\\
   \text{case 2} &\left\{ \begin{aligned} & M_1=-100~\text{GeV},~\tan\beta=60,\\
     & m_{L1}=m_{E2}=\frac{m_{E1}}{5}=\frac{m_{L2}}{10}=145~\text{GeV};
     \end{aligned}\right.
   \label{eq:ML145}
\end{eqnarray}
where the 45 GeV increment of $M_{L1}$ and $M_{E2}$ in Eq.~\eqref{eq:ML145} is to avoid the decay $\chi_2^0\to \ell\tilde{\ell}$ for escaping the CMS constraints~\cite{CMS:2021cox}. At the same time, the 45 GeV increment of $M_{L1}$ and $M_{E2}$ will not make $\delta a_{e,\,\mu}^{\rm{SUSY}}$ too small.

\begin{figure*}[tbp]
\centering 
\includegraphics[width=.4\textwidth]{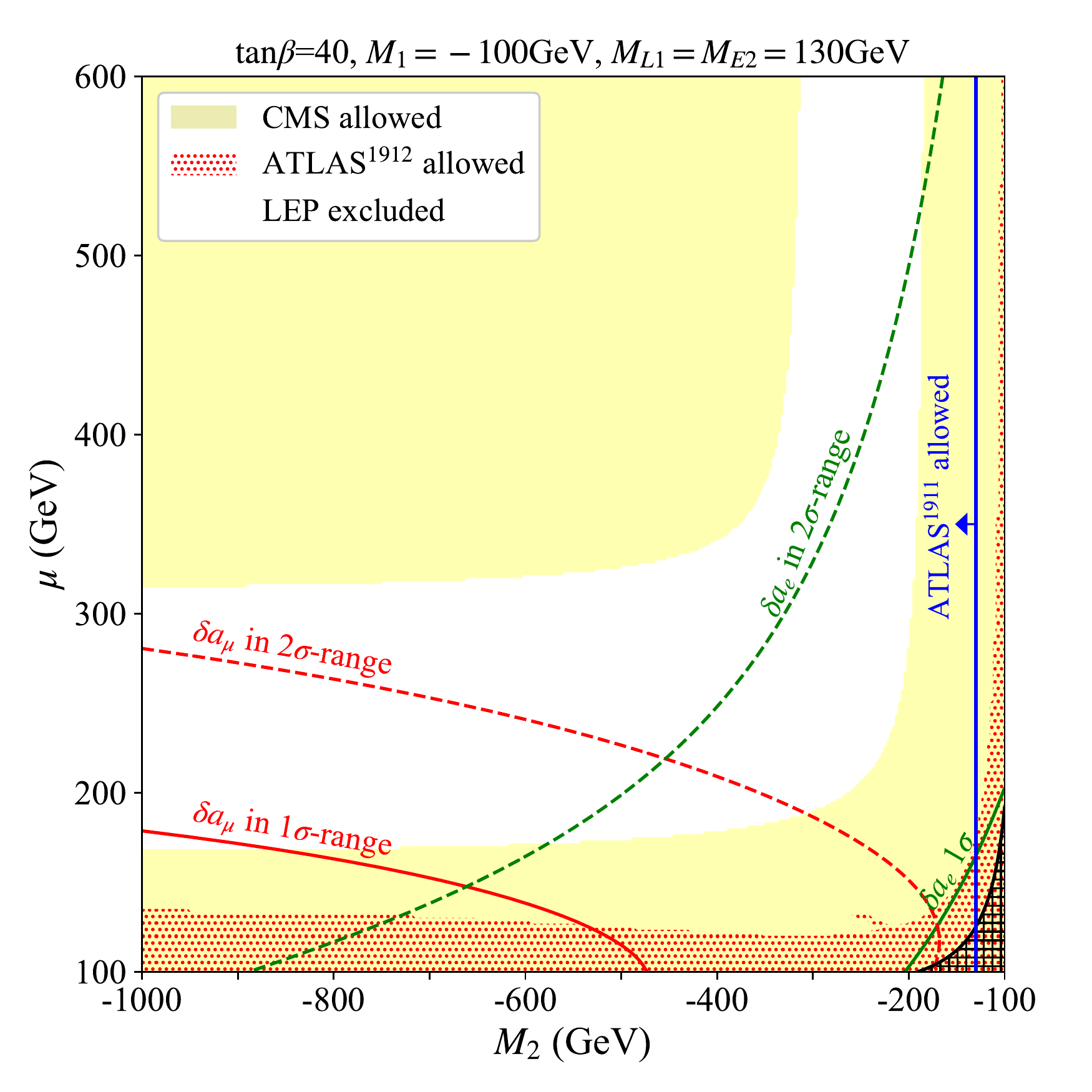}
\includegraphics[width=.4\textwidth]{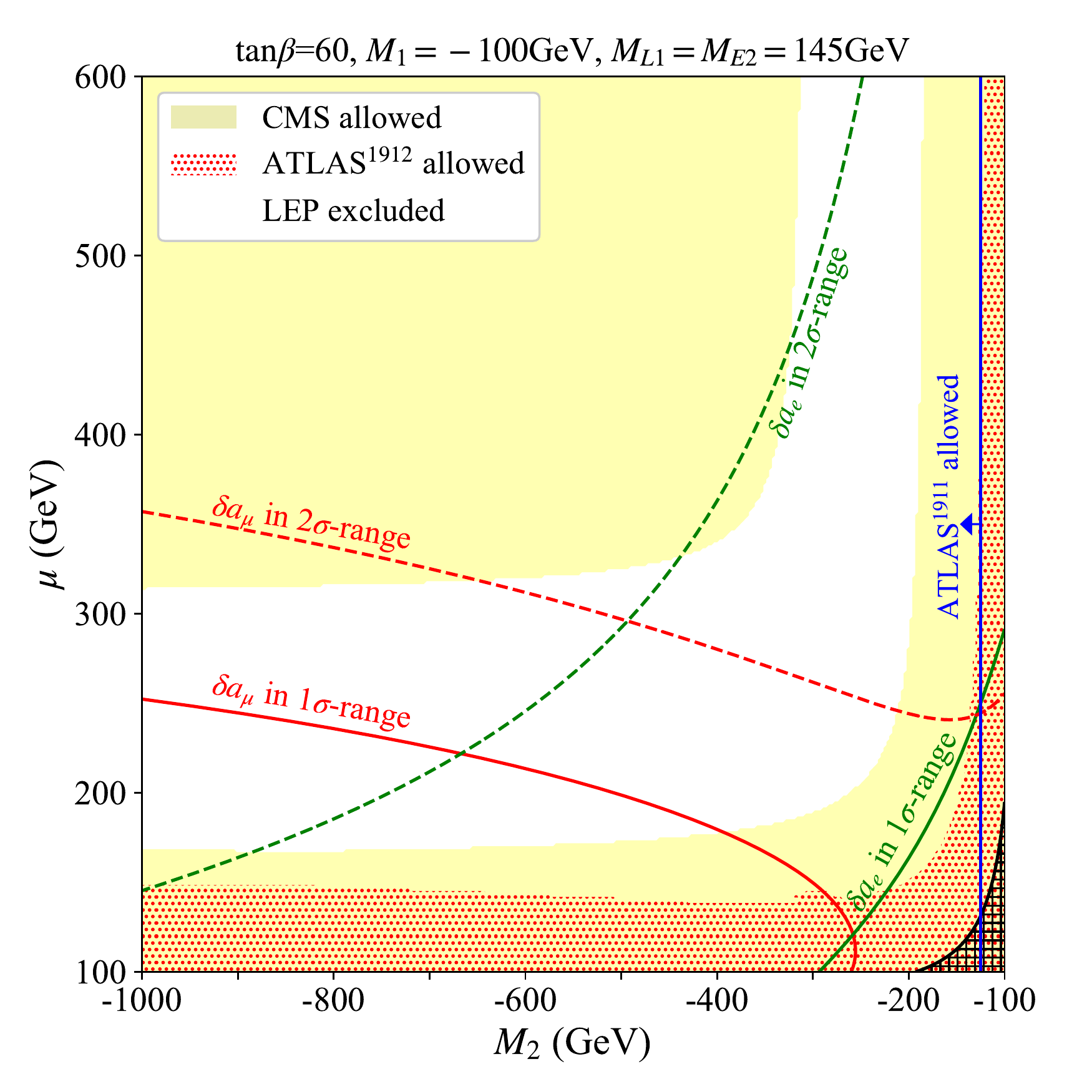}
\caption{\label{fig:LHCandGM2} The MSSM parameter space for a joint explanation of muon/electron $g-2$,  showing the LHC and LEP constraints. The constraints are from ATLAS1912~\cite{ATLAS:2019lng},  ATLAS1911~\cite{ATLAS:2019wgx}, CMS~\cite{CMS:2021cox} and  LEP~\cite{LEP:sleptons}. Other
constraints from the LHC ~\cite{ATLAS:2019lff,CMS:2020wxd} are not shown because they give similar results or have been considered in our scan (for an extensive recasting of LHC constraints, see, e.g., \cite{Athron:2021iuf}). For the ATLAS1911 limits, the regions indicated by the arrows are the allowed regions.} 
\end{figure*}

For these two cases we plot $\delta a_{e,\mu}^{\rm{SUSY}}$ in Fig.~\ref{fig:LHCandGM2}. 
We see that a joint explanation of $\Delta a_{e,\mu}^{\rm{Exp-SM}}$ may need a relatively 
a compressed spectrum for the bino-like $\chi^0_1$, the higgsino-like $\chi^0_2$ and $\chi^\pm_1$.    
For such compressed electroweakinos, the LHC performed the searches and gave the bounds ~\cite{ATLAS:2019lng}. 
Together with other searches for electroweakinos and sleptons at the LHC ~\cite{ATLAS:2019wgx,CMS:2021cox} and LEP ~\cite{LEP:sleptons}, the relevant experimental constraints are displayed in Fig.~\ref{fig:LHCandGM2}.   

From  Fig.~\ref{fig:LHCandGM2} we see that in both cases the results of the CMS Collaboration give strong limits on $\mu$: $\mu<130~\text{GeV}$ for $\tan\beta=40$ and $\mu<145\text{ GeV}$ for $\tan\beta=60$. 
For $\tan\beta=40$, $470~\text{GeV}<M_2<900~\text{GeV}$ is required for $\delta a_\mu^{\rm{SUSY}}$ within the $1\sigma$ range. However, the value of $M_2$ has a much wider range for $\tan\beta=60$. Of course, the constraints plotted in Fig.~\ref{fig:LHCandGM2} can be relaxed if the relevant decay branch ratios are not assumed to be $100\%$. So from Fig.~\ref{fig:LHCandGM2} we see that there indeed exist a MSSM parameter space  for a joint explanation of muon/electron $g-2$ anomalies.  

Note that here we used the constraints from the LHC searches in which the LSP is assumed to be the lightest neutralino.  
If the LSP is a superWIMP, the signals of the searched processes could be different, depending on the lifetime of the lightest neutralino. If the lightest neutralino has a relatively long lifetime and decays outside the detector ~\cite{Feng:2004gg}, the above LHC search constraints are applicable. If the lightest neutralino has a relatively short lifetime and decays inside the detector \cite{Liu:2014lda,Hikasa:2014yra}, the signals of the relevant processes will be different. In the latter case, the signals may be more
difficult to detect, for example, if the decay is dominated by $\chi^0_1\to \tilde G +h$ rather than by 
 $\chi^0_1\to \tilde G +Z/\gamma$ \cite{Liu:2014lda,Hikasa:2014yra}. 

Finally we should remark that in SUSY only the low energy effective MSSM has enough free parameters to possibly allow for a joint explanation of muon/electron $g-2$ anomalies (we do not go beyond the minimal framework of SUSY, albeit some extensions like NMSSM has the virtue of smaller fine-tuning extent confronting with the requirement of a 125 GeV Higgs boson, see, e.g.,  \cite{Cao:2012fz}, which can explain muon $g-2$ plus the AMS-02 anti-proton excess \cite{Abdughani:2021pdc}).  
The GUT-constrained models like mSUGRA cannot even explain the single anomaly of muon $g-2$
(the MSSM can readily explain the single anomaly of the muon $g-2$, see, e.g., \cite{Abdughani:2019wai,Cox:2018qyi,Kobakhidze:2016mdx,VanBeekveld:2021tgn,Baum:2021qzx}). 
For this end, some extensions have been proposed for these models, e.g., in  \cite{Akula:2013ioa,Wang:2015nra,Wang:2015rli,Wang:2017vxj,Wang:2018vrr,Han:2020exx,Li:2021pnt,Wang:2021bcx,Chakraborti:2021bmv}. 

\section{Conclusions}
\label{sec:conclusions}
Given the FNAL+BNL measurements for muon $g-2$ and 
the Berkeley $^{133}$Cs measurement for electron $g-2$, we explored the parameter space for a joint explanation, which requires a positive contribution to muon $g-2$ and a negative contribution to  electron $g-2$. Assuming no universality between smuon and selectron soft masses, we found out a part of parameter space for such a joint explanation at $2\sigma$ level, i.e., $\mu M_1,\mu M_2<0$, the masses of left selectron and right smuon below 200 GeV,  $m_{L2}$ much larger than the soft masses of other sleptons, $|M_1|<125$ GeV and $\mu<400$ GeV ($|M_2|$ is not subject to strict restrictions).
This part of parameter space can survive the LHC and LEP constraints, but gives an over-abundance for the dark matter if the bino-like lightest neutralino  is assumed to be the dark matter candidate. Then with the assumption that the dark matter candidate is a superWIMP (such as a pseudo-goldstino in multi-sector SUSY breaking scenarios, whose mass can be as light as GeV and produced from the late-decay of the thermally freeze-out lightest neutralinos), the dark matter problem can be avoided.  
So, we conclude that the MSSM may give a joint explanation for the muon and electron $g-2$ anomalies at $2\sigma$ level (the muon $g-2$ anomaly can be ameliorated to $1\sigma$).

\addcontentsline{toc}{section}{Acknowledgments}
\section*{Acknowledgments}
We thank Yang Zhang for helpful discussions. This work was supported by the National Natural Science Foundation of China (NNSFC) under grant Nos. 11821505 and 12075300,  
by Peng-Huan-Wu Theoretical Physics Innovation Center (12047503),
by the CAS Center for Excellence in Particle Physics (CCEPP), 
by the CAS Key Research Program of Frontier Sciences, 
and by a Key R\&D Program of Ministry of Science and Technology of China
under number 2017YFA0402204, and by the Key Research Program of the Chinese Academy of Sciences, Grant NO. XDPB15.

\addcontentsline{toc}{section}{Appendix}
\section*{Appendix}
The definitions of $n^{L,R}$, $c^{L,R}$ and the kinematic loop functions $F_{1,2}^{C,N}$ used in eq.\eqref{eq:leptonMDM1} and eq.\eqref{eq:leptonMDM2} are given by \cite{Martin:2001st}
\begin{align}
      \label{eq:nRim}
      n^R_{im} &= \sqrt2 g_1 N_{i1}X_{m2}^{(\ell)}+y_\ell N_{i3}X_{m1}^{(\ell)},\\
      \label{eq:nLim}
      n^L_{im} &= \frac{1}{\sqrt2}(g_2 N_{i2}+g_1 N_{i1})X^{(\ell)*}_{m1}-y_\ell N_{i3}X^{(\ell)*}_{m2},\\
      c^R_k &= y_\ell U_{k2},\\
      c^L_k &= -g_2 V_{k1},\\
      F_1^N(x) &= \frac{2}{(1-x)^4}\left(1-6x+3x^2+2x^3-6x^2\ln x\right),\\
      F_2^N(x) &= \frac{3}{(1-x)^3}\left(1-x^2+2x\ln x\right),\\
      F_1^C(x) &= \frac{2}{(1-x)^4}\left(2+3x-6x^2+x^3+6x\ln x\right),\\
      F_2^C(x)  &= -\frac{3}{2(1-x)^3}\left(3-4x+x^2+2\ln x\right),
\end{align}
where $y_\ell=g_2 m_\ell/(\sqrt2 m_W\cos\beta)$. $N$, $(U,V)$ and $X^{(\ell)}$ are the mixing matrices for the neutralinos, charginos and sleptons, respectively. In other words, these matrices satisfy
\begin{align}
    N^*M_{\chi^0}N^\dagger &= {\rm diag}(m_{\chi^0_1},m_{\chi^0_2},m_{\chi^0_3},m_{\chi^0_4}),\\
    U^*M_{\chi^\pm} V^\dagger &= {\rm diag}(m_{\chi^\pm_1},m_{\chi^\pm_2}), \\
    X^{(\ell)} M^2_{\tilde\ell} X^{(\ell)\dagger} &= {\rm diag}(m^2_{\tilde\ell_1}, m^2_{\tilde\ell_2}).
\end{align}

\addcontentsline{toc}{section}{References}
\bibliographystyle{JHEP}
\bibliography{bibliography}

\providecommand{\href}[2]{#2}\begingroup\raggedright\begin{thebibliography}{10}

\bibitem{PhysRevLett.126.141801}
{\scshape Muon $g\ensuremath{-}2$ Collaboration} collaboration, B.~Abi,
  T.~Albahri, S.~Al-Kilani et~al., \emph{Measurement of the positive muon
  anomalous magnetic moment to 0.46 ppm},
  \href{https://doi.org/10.1103/PhysRevLett.126.141801}{\emph{Phys. Rev. Lett.}
  {\bfseries 126} (Apr, 2021) 141801}.

\bibitem{Aoyama:2020ynm}
T.~Aoyama et~al., \emph{{The anomalous magnetic moment of the muon in the
  Standard Model}},
  \href{https://doi.org/10.1016/j.physrep.2020.07.006}{\emph{Phys. Rept.}
  {\bfseries 887} (2020) 1--166},
  [\href{https://arxiv.org/abs/2006.04822}{{\ttfamily 2006.04822}}].

\bibitem{Aoyama:2019ryr}
T.~Aoyama, T.~Kinoshita and M.~Nio, \emph{{Theory of the Anomalous Magnetic
  Moment of the Electron}},
  \href{https://doi.org/10.3390/atoms7010028}{\emph{Atoms} {\bfseries 7} (2019)
  28}.

\bibitem{Parker:2018vye}
R.~H. Parker, C.~Yu, W.~Zhong, B.~Estey and H.~M\"uller, \emph{{Measurement of
  the fine-structure constant as a test of the Standard Model}},
  \href{https://doi.org/10.1126/science.aap7706}{\emph{Science} {\bfseries 360}
  (2018) 191}, [\href{https://arxiv.org/abs/1812.04130}{{\ttfamily
  1812.04130}}].

\bibitem{Hanneke:2008tm}
D.~Hanneke, S.~Fogwell and G.~Gabrielse, \emph{{New Measurement of the Electron
  Magnetic Moment and the Fine Structure Constant}},
  \href{https://doi.org/10.1103/PhysRevLett.100.120801}{\emph{Phys. Rev. Lett.}
  {\bfseries 100} (2008) 120801},
  [\href{https://arxiv.org/abs/0801.1134}{{\ttfamily 0801.1134}}].

\bibitem{Morel:2020dww}
L.~Morel, Z.~Yao, P.~Clad\'e and S.~Guellati-Kh\'elifa, \emph{{Determination of
  the fine-structure constant with an accuracy of 81 parts per trillion}},
  \href{https://doi.org/10.1038/s41586-020-2964-7}{\emph{Nature} {\bfseries
  588} (2020) 61--65}.

\bibitem{Aoyama:2012wj}
T.~Aoyama, M.~Hayakawa, T.~Kinoshita and M.~Nio, \emph{{Tenth-Order QED
  Contribution to the Electron g-2 and an Improved Value of the Fine Structure
  Constant}}, \href{https://doi.org/10.1103/PhysRevLett.109.111807}{\emph{Phys.
  Rev. Lett.} {\bfseries 109} (2012) 111807},
  [\href{https://arxiv.org/abs/1205.5368}{{\ttfamily 1205.5368}}].

\bibitem{book:Jegerlehner:2017}
F.~Jegerlehner, \emph{The Anomalous Magnetic Moment of the Muon}.
\newblock Springer Tracts in Modern Physics volume 274. Springer, second
  edition~ed., 2017,
  \href{https://doi.org/10.1007/978-3-319-63577-4}{10.1007/978-3-319-63577-4}.

\bibitem{Giudice:2012ms}
G.~F. Giudice, P.~Paradisi and M.~Passera, \emph{{Testing new physics with the
  electron g-2}}, \href{https://doi.org/10.1007/JHEP11(2012)113}{\emph{JHEP}
  {\bfseries 11} (2012) 113},
  [\href{https://arxiv.org/abs/1208.6583}{{\ttfamily 1208.6583}}].

\bibitem{Hiller:2019mou}
G.~Hiller, C.~Hormigos-Feliu, D.~F. Litim and T.~Steudtner, \emph{{Anomalous
  magnetic moments from asymptotic safety}},
  \href{https://doi.org/10.1103/PhysRevD.102.071901}{\emph{Phys. Rev. D}
  {\bfseries 102} (2020) 071901},
  [\href{https://arxiv.org/abs/1910.14062}{{\ttfamily 1910.14062}}].

\bibitem{Hiller:2020fbu}
G.~Hiller, C.~Hormigos-Feliu, D.~F. Litim and T.~Steudtner, \emph{{Model
  Building from Asymptotic Safety with Higgs and Flavor Portals}},
  \href{https://doi.org/10.1103/PhysRevD.102.095023}{\emph{Phys. Rev. D}
  {\bfseries 102} (2020) 095023},
  [\href{https://arxiv.org/abs/2008.08606}{{\ttfamily 2008.08606}}].

\bibitem{Bissmann:2020lge}
S.~Bi\ss{}mann, G.~Hiller, C.~Hormigos-Feliu and D.~F. Litim,
  \emph{{Multi-lepton signatures of vector-like leptons with flavor}},
  \href{https://doi.org/10.1140/epjc/s10052-021-08886-3}{\emph{Eur. Phys. J. C}
  {\bfseries 81} (2021) 101},
  [\href{https://arxiv.org/abs/2011.12964}{{\ttfamily 2011.12964}}].

\bibitem{Rose:2020nxm}
L.~Delle~Rose, S.~Khalil and S.~Moretti, \emph{{Explaining electron and muon
  $g$ \ensuremath{-} 2 anomalies in an Aligned 2-Higgs Doublet Model with
  right-handed neutrinos}},
  \href{https://doi.org/10.1016/j.physletb.2021.136216}{\emph{Phys. Lett. B}
  {\bfseries 816} (2021) 136216},
  [\href{https://arxiv.org/abs/2012.06911}{{\ttfamily 2012.06911}}].

\bibitem{Botella:2020xzf}
F.~J. Botella, F.~Cornet-Gomez and M.~Nebot, \emph{{Electron and muon $g-2$
  anomalies in general flavour conserving two Higgs doublets models}},
  \href{https://doi.org/10.1103/PhysRevD.102.035023}{\emph{Phys. Rev. D}
  {\bfseries 102} (2020) 035023},
  [\href{https://arxiv.org/abs/2006.01934}{{\ttfamily 2006.01934}}].

\bibitem{Hernandez:2021tii}
A.~E.~C. Hern\'andez, S.~F. King and H.~Lee, \emph{{Fermion mass hierarchies
  from vector-like families with an extended 2HDM and a possible explanation
  for the electron and muon anomalous magnetic moments}},
  \href{https://arxiv.org/abs/2101.05819}{{\ttfamily 2101.05819}}.

\bibitem{Jana:2020pxx}
S.~Jana, V.~P. K. and S.~Saad, \emph{{Resolving electron and muon $g-2$ within
  the 2HDM}}, \href{https://doi.org/10.1103/PhysRevD.101.115037}{\emph{Phys.
  Rev. D} {\bfseries 101} (2020) 115037},
  [\href{https://arxiv.org/abs/2003.03386}{{\ttfamily 2003.03386}}].

\bibitem{Keung:2021rps}
W.-Y. Keung, D.~Marfatia and P.-Y. Tseng, \emph{{Axion-like particles,
  two-Higgs-doublet models, leptoquarks, and the electron and muon $g-2$}},
  \href{https://arxiv.org/abs/2104.03341}{{\ttfamily 2104.03341}}.

\bibitem{Li:2020dbg}
S.-P. Li, X.-Q. Li, Y.-Y. Li, Y.-D. Yang and X.~Zhang, \emph{{Power-aligned
  2HDM: a correlative perspective on $(g-2)_{e,\mu}$}},
  \href{https://doi.org/10.1007/JHEP01(2021)034}{\emph{JHEP} {\bfseries 01}
  (2021) 034}, [\href{https://arxiv.org/abs/2010.02799}{{\ttfamily
  2010.02799}}].

\bibitem{Han:2018znu}
X.-F. Han, T.~Li, L.~Wang and Y.~Zhang, \emph{{Simple interpretations of lepton
  anomalies in the lepton-specific inert two-Higgs-doublet model}},
  \href{https://doi.org/10.1103/PhysRevD.99.095034}{\emph{Phys. Rev. D}
  {\bfseries 99} (2019) 095034},
  [\href{https://arxiv.org/abs/1812.02449}{{\ttfamily 1812.02449}}].

\bibitem{Dorsner:2020aaz}
I.~Dor\v{s}ner, S.~Fajfer and S.~Saad, \emph{{$\mu \to e \gamma$ selecting
  scalar leptoquark solutions for the $(g-2)_{e,\mu}$ puzzles}},
  \href{https://doi.org/10.1103/PhysRevD.102.075007}{\emph{Phys. Rev. D}
  {\bfseries 102} (2020) 075007},
  [\href{https://arxiv.org/abs/2006.11624}{{\ttfamily 2006.11624}}].

\bibitem{Cornella:2019uxs}
C.~Cornella, P.~Paradisi and O.~Sumensari, \emph{{Hunting for ALPs with Lepton
  Flavor Violation}},
  \href{https://doi.org/10.1007/JHEP01(2020)158}{\emph{JHEP} {\bfseries 01}
  (2020) 158}, [\href{https://arxiv.org/abs/1911.06279}{{\ttfamily
  1911.06279}}].

\bibitem{Calibbi:2020emz}
L.~Calibbi, M.~L. L\'opez-Ib\'a\~nez, A.~Melis and O.~Vives, \emph{{Muon and
  electron $g-2$ and lepton masses in flavor models}},
  \href{https://doi.org/10.1007/JHEP06(2020)087}{\emph{JHEP} {\bfseries 06}
  (2020) 087}, [\href{https://arxiv.org/abs/2003.06633}{{\ttfamily
  2003.06633}}].

\bibitem{Escribano:2021css}
P.~Escribano, J.~Terol-Calvo and A.~Vicente,
  \emph{{$\boldsymbol{(g-2)_{e,\mu}}$ in an extended inverse type-III seesaw
  model}}, \href{https://doi.org/10.1103/PhysRevD.103.115018}{\emph{Phys. Rev.
  D} {\bfseries 103} (2021) 115018},
  [\href{https://arxiv.org/abs/2104.03705}{{\ttfamily 2104.03705}}].

\bibitem{Crivellin:2018qmi}
A.~Crivellin, M.~Hoferichter and P.~Schmidt-Wellenburg, \emph{{Combined
  explanations of $(g-2)_{\mu,e}$ and implications for a large muon EDM}},
  \href{https://doi.org/10.1103/PhysRevD.98.113002}{\emph{Phys. Rev. D}
  {\bfseries 98} (2018) 113002},
  [\href{https://arxiv.org/abs/1807.11484}{{\ttfamily 1807.11484}}].

\bibitem{TheMEG:2016wtm}
{\scshape MEG} collaboration, A.~M. Baldini et~al., \emph{{Search for the
  lepton flavour violating decay $\mu ^+ \rightarrow \mathrm {e}^+ \gamma $
  with the full dataset of the MEG experiment}},
  \href{https://doi.org/10.1140/epjc/s10052-016-4271-x}{\emph{Eur. Phys. J. C}
  {\bfseries 76} (2016) 434},
  [\href{https://arxiv.org/abs/1605.05081}{{\ttfamily 1605.05081}}].

\bibitem{CarcamoHernandez:2019ydc}
A.~E. C\'arcamo~Hern\'andez, S.~F. King, H.~Lee and S.~J. Rowley, \emph{{Is it
  possible to explain the muon and electron $g-2$ in a $Z'$ model?}},
  \href{https://doi.org/10.1103/PhysRevD.101.115016}{\emph{Phys. Rev. D}
  {\bfseries 101} (2020) 115016},
  [\href{https://arxiv.org/abs/1910.10734}{{\ttfamily 1910.10734}}].

\bibitem{Bodas:2021fsy}
A.~Bodas, R.~Coy and S.~J.~D. King, \emph{{Solving the electron and muon $g-2$
  anomalies in $Z'$ models}},
  \href{https://arxiv.org/abs/2102.07781}{{\ttfamily 2102.07781}}.

\bibitem{Chen:2020tfr}
K.-F. Chen, C.-W. Chiang and K.~Yagyu, \emph{{An explanation for the muon and
  electron $g-2$ anomalies and dark matter}},
  \href{https://doi.org/10.1007/JHEP09(2020)119}{\emph{JHEP} {\bfseries 09}
  (2020) 119}, [\href{https://arxiv.org/abs/2006.07929}{{\ttfamily
  2006.07929}}].

\bibitem{Dutta:2018fge}
B.~Dutta and Y.~Mimura, \emph{{Electron $g-2$ with flavor violation in MSSM}},
  \href{https://doi.org/10.1016/j.physletb.2018.12.070}{\emph{Phys. Lett. B}
  {\bfseries 790} (2019) 563--567},
  [\href{https://arxiv.org/abs/1811.10209}{{\ttfamily 1811.10209}}].

\bibitem{Badziak:2019gaf}
M.~Badziak and K.~Sakurai, \emph{{Explanation of electron and muon g
  \ensuremath{-} 2 anomalies in the MSSM}},
  \href{https://doi.org/10.1007/JHEP10(2019)024}{\emph{JHEP} {\bfseries 10}
  (2019) 024}, [\href{https://arxiv.org/abs/1908.03607}{{\ttfamily
  1908.03607}}].

\bibitem{Endo:2019bcj}
M.~Endo and W.~Yin, \emph{{Explaining electron and muon $g-2$ anomaly in SUSY
  without lepton-flavor mixings}},
  \href{https://doi.org/10.1007/JHEP08(2019)122}{\emph{JHEP} {\bfseries 08}
  (2019) 122}, [\href{https://arxiv.org/abs/1906.08768}{{\ttfamily
  1906.08768}}].

\bibitem{Yang:2020bmh}
J.-L. Yang, T.-F. Feng and H.-B. Zhang, \emph{{Electron and muon $(g-2)$ in the
  B-LSSM}}, \href{https://doi.org/10.1088/1361-6471/ab7986}{\emph{J. Phys. G}
  {\bfseries 47} (2020) 055004},
  [\href{https://arxiv.org/abs/2003.09781}{{\ttfamily 2003.09781}}].

\bibitem{Cao:2021lmj}
J.~Cao, Y.~He, J.~Lian, D.~Zhang and P.~Zhu, \emph{{Electron and Muon Anomalous
  Magnetic Moments in the Inverse Seesaw Extended NMSSM}},
  \href{https://arxiv.org/abs/2102.11355}{{\ttfamily 2102.11355}}.

\bibitem{Martin:2001st}
S.~P. Martin and J.~D. Wells, \emph{{Muon Anomalous Magnetic Dipole Moment in
  Supersymmetric Theories}},
  \href{https://doi.org/10.1103/PhysRevD.64.035003}{\emph{Phys. Rev. D}
  {\bfseries 64} (2001) 035003},
  [\href{https://arxiv.org/abs/hep-ph/0103067}{{\ttfamily hep-ph/0103067}}].

\bibitem{Degrassi:1998es}
G.~Degrassi and G.~F. Giudice, \emph{{QED logarithms in the electroweak
  corrections to the muon anomalous magnetic moment}},
  \href{https://doi.org/10.1103/PhysRevD.58.053007}{\emph{Phys. Rev. D}
  {\bfseries 58} (1998) 053007},
  [\href{https://arxiv.org/abs/hep-ph/9803384}{{\ttfamily hep-ph/9803384}}].

\bibitem{Marchetti:2008hw}
S.~Marchetti, S.~Mertens, U.~Nierste and D.~Stockinger,
  \emph{{Tan(beta)-enhanced supersymmetric corrections to the anomalous
  magnetic moment of the muon}},
  \href{https://doi.org/10.1103/PhysRevD.79.013010}{\emph{Phys. Rev. D}
  {\bfseries 79} (2009) 013010},
  [\href{https://arxiv.org/abs/0808.1530}{{\ttfamily 0808.1530}}].

\bibitem{Carena:1999py}
M.~Carena, D.~Garcia, U.~Nierste and C.~E.~M. Wagner, \emph{{Effective
  Lagrangian for the $\bar{t} b H^{+}$ interaction in the MSSM and charged
  Higgs phenomenology}},
  \href{https://doi.org/10.1016/S0550-3213(00)00146-2}{\emph{Nucl. Phys. B}
  {\bfseries 577} (2000) 88--120},
  [\href{https://arxiv.org/abs/hep-ph/9912516}{{\ttfamily hep-ph/9912516}}].

\bibitem{Lee:2003nta}
J.~S. Lee, A.~Pilaftsis, M.~Carena, S.~Y. Choi, M.~Drees, J.~R. Ellis et~al.,
  \emph{{CPsuperH: A Computational tool for Higgs phenomenology in the minimal
  supersymmetric standard model with explicit CP violation}},
  \href{https://doi.org/10.1016/S0010-4655(03)00463-6}{\emph{Comput. Phys.
  Commun.} {\bfseries 156} (2004) 283--317},
  [\href{https://arxiv.org/abs/hep-ph/0307377}{{\ttfamily hep-ph/0307377}}].

\bibitem{Lee:2007gn}
J.~S. Lee, M.~Carena, J.~Ellis, A.~Pilaftsis and C.~E.~M. Wagner,
  \emph{{CPsuperH2.0: an Improved Computational Tool for Higgs Phenomenology in
  the MSSM with Explicit CP Violation}},
  \href{https://doi.org/10.1016/j.cpc.2008.09.003}{\emph{Comput. Phys. Commun.}
  {\bfseries 180} (2009) 312--331},
  [\href{https://arxiv.org/abs/0712.2360}{{\ttfamily 0712.2360}}].

\bibitem{Lee:2012wa}
J.~S. Lee, M.~Carena, J.~Ellis, A.~Pilaftsis and C.~E.~M. Wagner,
  \emph{{CPsuperH2.3: an Updated Tool for Phenomenology in the MSSM with
  Explicit CP Violation}},
  \href{https://doi.org/10.1016/j.cpc.2012.11.006}{\emph{Comput. Phys. Commun.}
  {\bfseries 184} (2013) 1220--1233},
  [\href{https://arxiv.org/abs/1208.2212}{{\ttfamily 1208.2212}}].

\bibitem{Moroi:1995yh}
T.~Moroi, \emph{{The Muon anomalous magnetic dipole moment in the minimal
  supersymmetric standard model}},
  \href{https://doi.org/10.1103/PhysRevD.53.6565}{\emph{Phys. Rev. D}
  {\bfseries 53} (1996) 6565--6575},
  [\href{https://arxiv.org/abs/hep-ph/9512396}{{\ttfamily hep-ph/9512396}}].

\bibitem{Stockinger:2006zn}
D.~Stockinger, \emph{{The Muon Magnetic Moment and Supersymmetry}},
  \href{https://doi.org/10.1088/0954-3899/34/2/R01}{\emph{J. Phys. G}
  {\bfseries 34} (2007) R45--R92},
  [\href{https://arxiv.org/abs/hep-ph/0609168}{{\ttfamily hep-ph/0609168}}].

\bibitem{LEP:sleptons}
{\scshape LEP2 SUSY Working Group} collaboration, ``Combined {LEP
  Selectron/Smuon/Stau Results}, 183-208{GeV}.''
  \url{http://lepsusy.web.cern.ch/lepsusy/www/sleptons_summer02/slep_2002.html},
  2002.

\bibitem{ATLAS:2019lng}
{\scshape ATLAS} collaboration, G.~Aad et~al., \emph{{Searches for electroweak
  production of supersymmetric particles with compressed mass spectra in
  $\sqrt{s}=$ 13 TeV $pp$ collisions with the ATLAS detector}},
  \href{https://doi.org/10.1103/PhysRevD.101.052005}{\emph{Phys. Rev. D}
  {\bfseries 101} (2020) 052005},
  [\href{https://arxiv.org/abs/1911.12606}{{\ttfamily 1911.12606}}].

\bibitem{Athron:2014yba}
P.~Athron, J.-h. Park, D.~St\"ockinger and A.~Voigt,
  \emph{{FlexibleSUSY\textemdash{}A spectrum generator generator for
  supersymmetric models}},
  \href{https://doi.org/10.1016/j.cpc.2014.12.020}{\emph{Comput. Phys. Commun.}
  {\bfseries 190} (2015) 139--172},
  [\href{https://arxiv.org/abs/1406.2319}{{\ttfamily 1406.2319}}].

\bibitem{Athron:2017fvs}
P.~Athron, M.~Bach, D.~Harries, T.~Kwasnitza, J.-h. Park, D.~St\"ockinger
  et~al., \emph{{FlexibleSUSY 2.0: Extensions to investigate the phenomenology
  of SUSY and non-SUSY models}},
  \href{https://doi.org/10.1016/j.cpc.2018.04.016}{\emph{Comput. Phys. Commun.}
  {\bfseries 230} (2018) 145--217},
  [\href{https://arxiv.org/abs/1710.03760}{{\ttfamily 1710.03760}}].

\bibitem{Belanger:2006is}
G.~Belanger, F.~Boudjema, A.~Pukhov and A.~Semenov, \emph{{MicrOMEGAs 2.0: A
  Program to calculate the relic density of dark matter in a generic model}},
  \href{https://doi.org/10.1016/j.cpc.2006.11.008}{\emph{Comput. Phys. Commun.}
  {\bfseries 176} (2007) 367--382},
  [\href{https://arxiv.org/abs/hep-ph/0607059}{{\ttfamily hep-ph/0607059}}].

\bibitem{Belanger:2008sj}
G.~Belanger, F.~Boudjema, A.~Pukhov and A.~Semenov, \emph{{Dark matter direct
  detection rate in a generic model with micrOMEGAs 2.2}},
  \href{https://doi.org/10.1016/j.cpc.2008.11.019}{\emph{Comput. Phys. Commun.}
  {\bfseries 180} (2009) 747--767},
  [\href{https://arxiv.org/abs/0803.2360}{{\ttfamily 0803.2360}}].

\bibitem{Belanger:2010pz}
G.~Belanger, F.~Boudjema, A.~Pukhov and A.~Semenov, \emph{{micrOMEGAs: A Tool
  for dark matter studies}},
  \href{https://doi.org/10.1393/ncc/i2010-10591-3}{\emph{Nuovo Cim. C}
  {\bfseries 033N2} (2010) 111--116},
  [\href{https://arxiv.org/abs/1005.4133}{{\ttfamily 1005.4133}}].

\bibitem{Belanger:2013oya}
G.~Belanger, F.~Boudjema, A.~Pukhov and A.~Semenov, \emph{{micrOMEGAs$\_$3: A
  program for calculating dark matter observables}},
  \href{https://doi.org/10.1016/j.cpc.2013.10.016}{\emph{Comput. Phys. Commun.}
  {\bfseries 185} (2014) 960--985},
  [\href{https://arxiv.org/abs/1305.0237}{{\ttfamily 1305.0237}}].

\bibitem{Planck:2018vyg}
{\scshape Planck} collaboration, N.~Aghanim et~al., \emph{{Planck 2018 results.
  VI. Cosmological parameters}},
  \href{https://doi.org/10.1051/0004-6361/201833910}{\emph{Astron. Astrophys.}
  {\bfseries 641} (2020) A6},
  [\href{https://arxiv.org/abs/1807.06209}{{\ttfamily 1807.06209}}].

\bibitem{XENON:2017vdw}
{\scshape XENON} collaboration, E.~Aprile et~al., \emph{{First Dark Matter
  Search Results from the XENON1T Experiment}},
  \href{https://doi.org/10.1103/PhysRevLett.119.181301}{\emph{Phys. Rev. Lett.}
  {\bfseries 119} (2017) 181301},
  [\href{https://arxiv.org/abs/1705.06655}{{\ttfamily 1705.06655}}].

\bibitem{Feng:2004gg}
J.~L. Feng, S.-f. Su and F.~Takayama, \emph{{SuperWIMP dark matter in
  supergravity with a gravitino LSP}},  in \emph{{12th International Conference
  on Supersymmetry and Unification of Fundamental Interactions (SUSY 04)}}, 10,
  2004, \href{https://arxiv.org/abs/hep-ph/0410119}{{\ttfamily
  hep-ph/0410119}}.

\bibitem{Wang:2004ib}
F.~Wang and J.~M. Yang, \emph{{SuperWIMP dark matter scenario in light of
  WMAP}}, \href{https://doi.org/10.1140/epjc/s2004-02029-6}{\emph{Eur. Phys. J.
  C} {\bfseries 38} (2004) 129--133},
  [\href{https://arxiv.org/abs/hep-ph/0405186}{{\ttfamily hep-ph/0405186}}].

\bibitem{Argurio:2011hs}
R.~Argurio, Z.~Komargodski and A.~Mariotti, \emph{{Pseudo-Goldstini in Field
  Theory}}, \href{https://doi.org/10.1103/PhysRevLett.107.061601}{\emph{Phys.
  Rev. Lett.} {\bfseries 107} (2011) 061601},
  [\href{https://arxiv.org/abs/1102.2386}{{\ttfamily 1102.2386}}].

\bibitem{Dai:2021eah}
J.~Dai, T.~Liu and J.~M. Yang, \emph{{An explicit calculation of
  pseudo-goldstino mass at the leading three-loop level}},
  \href{https://doi.org/10.1007/JHEP06(2021)175}{\emph{JHEP} {\bfseries 06}
  (2021) 175}, [\href{https://arxiv.org/abs/2104.12656}{{\ttfamily
  2104.12656}}].

\bibitem{Cyburt:2002uv}
R.~H. Cyburt, J.~R. Ellis, B.~D. Fields and K.~A. Olive, \emph{{Updated
  nucleosynthesis constraints on unstable relic particles}},
  \href{https://doi.org/10.1103/PhysRevD.67.103521}{\emph{Phys. Rev. D}
  {\bfseries 67} (2003) 103521},
  [\href{https://arxiv.org/abs/astro-ph/0211258}{{\ttfamily
  astro-ph/0211258}}].

\bibitem{Gu:2020ozv}
Y.~Gu, M.~Khlopov, L.~Wu, J.~M. Yang and B.~Zhu, \emph{{Light gravitino dark
  matter: LHC searches and the Hubble tension}},
  \href{https://doi.org/10.1103/PhysRevD.102.115005}{\emph{Phys. Rev. D}
  {\bfseries 102} (2020) 115005},
  [\href{https://arxiv.org/abs/2006.09906}{{\ttfamily 2006.09906}}].

\bibitem{Chen:2021omv}
J.~Chen, C.~Han, J.~M. Yang and M.~Zhang, \emph{{Probing a bino NLSP at lepton
  colliders}}, \href{https://doi.org/10.1103/PhysRevD.104.015009}{\emph{Phys.
  Rev. D} {\bfseries 104} (2021) 015009},
  [\href{https://arxiv.org/abs/2101.12131}{{\ttfamily 2101.12131}}].

\bibitem{Porod:2003um}
W.~Porod, \emph{{SPheno, a program for calculating supersymmetric spectra, SUSY
  particle decays and SUSY particle production at e+ e- colliders}},
  \href{https://doi.org/10.1016/S0010-4655(03)00222-4}{\emph{Comput. Phys.
  Commun.} {\bfseries 153} (2003) 275--315},
  [\href{https://arxiv.org/abs/hep-ph/0301101}{{\ttfamily hep-ph/0301101}}].

\bibitem{Porod:2011nf}
W.~Porod and F.~Staub, \emph{{SPheno 3.1: Extensions including flavour,
  CP-phases and models beyond the MSSM}},
  \href{https://doi.org/10.1016/j.cpc.2012.05.021}{\emph{Comput. Phys. Commun.}
  {\bfseries 183} (2012) 2458--2469},
  [\href{https://arxiv.org/abs/1104.1573}{{\ttfamily 1104.1573}}].

\bibitem{CMS:2021cox}
{\scshape CMS} collaboration, A.~Tumasyan et~al., \emph{{Search for electroweak
  production of charginos and neutralinos in proton-proton collisions at
  $\sqrt{s} = $ 13 TeV}},  \href{https://arxiv.org/abs/2106.14246}{{\ttfamily
  2106.14246}}.

\bibitem{ATLAS:2019wgx}
{\scshape ATLAS} collaboration, G.~Aad et~al., \emph{{Search for
  chargino-neutralino production with mass splittings near the electroweak
  scale in three-lepton final states in $\sqrt {s}$=13 TeV $pp$ collisions with
  the ATLAS detector}},
  \href{https://doi.org/10.1103/PhysRevD.101.072001}{\emph{Phys. Rev. D}
  {\bfseries 101} (2020) 072001},
  [\href{https://arxiv.org/abs/1912.08479}{{\ttfamily 1912.08479}}].

\bibitem{ATLAS:2019lff}
{\scshape ATLAS} collaboration, G.~Aad et~al., \emph{{Search for electroweak
  production of charginos and sleptons decaying into final states with two
  leptons and missing transverse momentum in $\sqrt{s}=13$ TeV $pp$ collisions
  using the ATLAS detector}},
  \href{https://doi.org/10.1140/epjc/s10052-019-7594-6}{\emph{Eur. Phys. J. C}
  {\bfseries 80} (2020) 123},
  [\href{https://arxiv.org/abs/1908.08215}{{\ttfamily 1908.08215}}].

\bibitem{CMS:2020wxd}
{\scshape CMS} collaboration, \emph{{Search for physics beyond the standard
  model in final states with two opposite-charge same-flavor leptons, jets, and
  missing transverse momentum in pp collisions at 13 TeV}}, .

\bibitem{Athron:2021iuf}
P.~Athron, C.~Bal\'azs, D.~H. Jacob, W.~Kotlarski, D.~St\"ockinger and
  H.~St\"ockinger-Kim, \emph{{New physics explanations of $a_\mu$ in light of
  the FNAL muon $g-2$ measurement}},
  \href{https://arxiv.org/abs/2104.03691}{{\ttfamily 2104.03691}}.

\bibitem{Liu:2014lda}
T.~Liu, L.~Wang and J.~M. Yang, \emph{{Pseudo-goldstino and electroweakinos via
  VBF processes at LHC}},
  \href{https://doi.org/10.1007/JHEP02(2015)177}{\emph{JHEP} {\bfseries 02}
  (2015) 177}, [\href{https://arxiv.org/abs/1411.6105}{{\ttfamily 1411.6105}}].

\bibitem{Hikasa:2014yra}
K.-i. Hikasa, T.~Liu, L.~Wang and J.~M. Yang, \emph{{Pseudo-goldstino and
  electroweak gauginos at the LHC}},
  \href{https://doi.org/10.1007/JHEP07(2014)065}{\emph{JHEP} {\bfseries 07}
  (2014) 065}, [\href{https://arxiv.org/abs/1403.5731}{{\ttfamily 1403.5731}}].

\bibitem{Cao:2012fz}
J.-J. Cao, Z.-X. Heng, J.~M. Yang, Y.-M. Zhang and J.-Y. Zhu, \emph{{A SM-like
  Higgs near 125 GeV in low energy SUSY: a comparative study for MSSM and
  NMSSM}}, \href{https://doi.org/10.1007/JHEP03(2012)086}{\emph{JHEP}
  {\bfseries 03} (2012) 086},
  [\href{https://arxiv.org/abs/1202.5821}{{\ttfamily 1202.5821}}].

\bibitem{Abdughani:2021pdc}
M.~Abdughani, Y.-Z. Fan, L.~Feng, Y.-L. Sming~Tsai, L.~Wu and Q.~Yuan, \emph{{A
  common origin of muon g-2 anomaly, Galaxy Center GeV excess and AMS-02
  anti-proton excess in the NMSSM}},
  \href{https://arxiv.org/abs/2104.03274}{{\ttfamily 2104.03274}}.

\bibitem{Abdughani:2019wai}
M.~Abdughani, K.-I. Hikasa, L.~Wu, J.~M. Yang and J.~Zhao, \emph{{Testing
  electroweak SUSY for muon $g$ \ensuremath{-} 2 and dark matter at the LHC and
  beyond}}, \href{https://doi.org/10.1007/JHEP11(2019)095}{\emph{JHEP}
  {\bfseries 11} (2019) 095},
  [\href{https://arxiv.org/abs/1909.07792}{{\ttfamily 1909.07792}}].

\bibitem{Cox:2018qyi}
P.~Cox, C.~Han and T.~T. Yanagida, \emph{{Muon $g-2$ and dark matter in the
  minimal supersymmetric standard model}},
  \href{https://doi.org/10.1103/PhysRevD.98.055015}{\emph{Phys. Rev. D}
  {\bfseries 98} (2018) 055015},
  [\href{https://arxiv.org/abs/1805.02802}{{\ttfamily 1805.02802}}].

\bibitem{Kobakhidze:2016mdx}
A.~Kobakhidze, M.~Talia and L.~Wu, \emph{{Probing the MSSM explanation of the
  muon g-2 anomaly in dark matter experiments and at a 100 TeV $pp$ collider}},
  \href{https://doi.org/10.1103/PhysRevD.95.055023}{\emph{Phys. Rev. D}
  {\bfseries 95} (2017) 055023},
  [\href{https://arxiv.org/abs/1608.03641}{{\ttfamily 1608.03641}}].

\bibitem{VanBeekveld:2021tgn}
M.~Van~Beekveld, W.~Beenakker, M.~Schutten and J.~De~Wit, \emph{{Dark matter,
  fine-tuning and $(g-2)_{\mu}$ in the pMSSM}},
  \href{https://arxiv.org/abs/2104.03245}{{\ttfamily 2104.03245}}.

\bibitem{Baum:2021qzx}
S.~Baum, M.~Carena, N.~R. Shah and C.~E.~M. Wagner, \emph{{The Tiny (g-2) Muon
  Wobble from Small-$\mu$ Supersymmetry}},
  \href{https://arxiv.org/abs/2104.03302}{{\ttfamily 2104.03302}}.

\bibitem{Akula:2013ioa}
S.~Akula and P.~Nath, \emph{{Gluino-driven radiative breaking, Higgs boson
  mass, muon g-2, and the Higgs diphoton decay in supergravity unification}},
  \href{https://doi.org/10.1103/PhysRevD.87.115022}{\emph{Phys. Rev. D}
  {\bfseries 87} (2013) 115022},
  [\href{https://arxiv.org/abs/1304.5526}{{\ttfamily 1304.5526}}].

\bibitem{Wang:2015nra}
F.~Wang, W.~Wang, J.~M. Yang and Y.~Zhang, \emph{{Heavy colored SUSY partners
  from deflected anomaly mediation}},
  \href{https://doi.org/10.1007/JHEP07(2015)138}{\emph{JHEP} {\bfseries 07}
  (2015) 138}, [\href{https://arxiv.org/abs/1505.02785}{{\ttfamily
  1505.02785}}].

\bibitem{Wang:2015rli}
F.~Wang, W.~Wang and J.~M. Yang, \emph{{Reconcile muon g-2 anomaly with LHC
  data in SUGRA with generalized gravity mediation}},
  \href{https://doi.org/10.1007/JHEP06(2015)079}{\emph{JHEP} {\bfseries 06}
  (2015) 079}, [\href{https://arxiv.org/abs/1504.00505}{{\ttfamily
  1504.00505}}].

\bibitem{Wang:2017vxj}
F.~Wang, W.~Wang and J.~M. Yang, \emph{{Solving the muon g-2 anomaly in
  deflected anomaly mediated SUSY breaking with messenger-matter
  interactions}}, \href{https://doi.org/10.1103/PhysRevD.96.075025}{\emph{Phys.
  Rev. D} {\bfseries 96} (2017) 075025},
  [\href{https://arxiv.org/abs/1703.10894}{{\ttfamily 1703.10894}}].

\bibitem{Wang:2018vrr}
F.~Wang, K.~Wang, J.~M. Yang and J.~Zhu, \emph{{Solving the muon g-2 anomaly in
  CMSSM extension with non-universal gaugino masses}},
  \href{https://doi.org/10.1007/JHEP12(2018)041}{\emph{JHEP} {\bfseries 12}
  (2018) 041}, [\href{https://arxiv.org/abs/1808.10851}{{\ttfamily
  1808.10851}}].

\bibitem{Han:2020exx}
C.~Han, M.~L. L\'opez-Ib\'a\~nez, A.~Melis, O.~Vives, L.~Wu and J.~M. Yang,
  \emph{{LFV and (g-2) in non-universal SUSY models with light higgsinos}},
  \href{https://doi.org/10.1007/JHEP05(2020)102}{\emph{JHEP} {\bfseries 05}
  (2020) 102}, [\href{https://arxiv.org/abs/2003.06187}{{\ttfamily
  2003.06187}}].

\bibitem{Li:2021pnt}
Z.~Li, G.-L. Liu, F.~Wang, J.~M. Yang and Y.~Zhang, \emph{{Gluino-SUGRA
  scenarios in light of FNAL muon g-2 anomaly}},
  \href{https://arxiv.org/abs/2106.04466}{{\ttfamily 2106.04466}}.

\bibitem{Wang:2021bcx}
F.~Wang, L.~Wu, Y.~Xiao, J.~M. Yang and Y.~Zhang, \emph{{GUT-scale constrained
  SUSY in light of E989 muon g-2 measurement}},
  \href{https://arxiv.org/abs/2104.03262}{{\ttfamily 2104.03262}}.

\bibitem{Chakraborti:2021bmv}
M.~Chakraborti, L.~Roszkowski and S.~Trojanowski, \emph{{GUT-constrained
  supersymmetry and dark matter in light of the new $(g-2)_\mu$
  determination}}, \href{https://doi.org/10.1007/JHEP05(2021)252}{\emph{JHEP}
  {\bfseries 05} (2021) 252},
  [\href{https://arxiv.org/abs/2104.04458}{{\ttfamily 2104.04458}}].

\end{thebibliography}\endgroup

\end{document}